\documentclass{llncs}

\usepackage[usenames]{color}
\usepackage{array}
\usepackage{graphicx}
\usepackage[all]{xy}
\xyoption{dvips}
\usepackage{amssymb,amsmath}
\usepackage[mathcal]{euscript}
\usepackage{proof}

\usepackage{paper}

\title{A linear-non-linear model for a computational call-by-value lambda calculus\\(extended abstract)}

\def\tablenl{\\[0ex]}
\newcommand{\mycaption}[1]{\caption{#1}\vspace{-5ex}}

\author{Peter Selinger\inst{1} \and Beno\^{\i}t Valiron\inst{2}}

\institute{
  Dalhousie University,
  \email{selinger@mathstat.dal.ca}
  \and
  University of Ottawa,
  \email{bvali087@uottawa.ca}
}

\begin{document}

\maketitle

\begin{abstract}
  We give a categorical semantics for a call-by-value linear lambda
  calculus. Such a lambda calculus was used by Selinger and Valiron as
  the backbone of a functional programming language for quantum
  computation. One feature of this lambda calculus is its linear type system,
  which includes a duplicability operator ``$!$'' as in linear logic.
  Another main feature is its call-by-value reduction strategy,
  together with a side-effect to model probabilistic measurements. The
  ``$!$'' operator gives rise to a comonad, as in the linear logic
  models of Seely, Bierman, and Benton. The side-effects give rise to
  a monad, as in Moggi's computational lambda calculus.  It is this
  combination of a monad and a comonad that makes the present paper
  interesting. We show that our categorical semantics is sound and
  complete.
\end{abstract}

\section{Introduction}

In the last few years, there has been some interest in the semantics
of quantum programming languages.
{\cite{selinger04quantum}} gave a denotational
semantics for a flow-chart language, but this language did not include
higher-order types. Several authors defined quantum lambda calculi
{\cite{tonder04lambda,selinger05lambda}} as well as quantum process
algebras {\cite{GaySJ:comqp,LalJor}}, which had higher-order features
and a well-defined operational semantics, but lacked denotational
semantics. \cite{valiron06fully} gave a categorical model for a
higher-order quantum lambda calculus, but omitted all the non-linear
features (i.e., classical data).  Meanwhile, Abramsky and Coecke
{\cite{abramsky04categorical,coecke04informationflow}}
developed categorical axiomatics for Hilbert spaces, but there is
no particular language associated with these models.

In this paper, we give the first categorical semantics of an
unabridged quantum lambda calculus, which is a version of the language
studied in {\cite{selinger05lambda}}. 

For the purposes of the present paper, an understanding of the precise
mechanics of quantum computation is not required. We will focus
primarily on the type system and language, and not on the structure of
the actual ``built-in'' quantum operations (such as unitary operators
and measurements). In this sense, this paper is about the semantics of
a generic call-by-value linear lambda calculus, which is parametric on
some primitive operations that are not further explained. It should be
understood, however, that the need to support primitive quantum
operations motivates particular features of the type system, which we
briefly explain now.

The first important language feature is linearity. This arises from
the well-known {\em no-cloning} property of quantum computation, which
asserts that quantum data cannot be duplicated {\cite{WZ82}}. So if
$x:\qbit$ is a variable representing a quantum bit, and $y:\bit$ is a
variable representing a classical bit, then it is legal to write
$f(y,y)$, but not $g(x,x)$. 
In order to keep track of duplicability at higher-order types we use
a type system based on linear logic. We use the duplicability operator
``$!$'' to mark classical types.
In the categorical semantics, this
operator gives rise to a comonad as in the work of {\cite{Seely89}}
and {\cite{benton92linear}}.
Another account of mixing copyable and non-copyable data is~\cite{coecke06quantum},
where the copyability is internal to objects.

A second feature of quantum computation is its probabilistic nature.
Quantum physics has an operation called measurement, which converts
quantum data to classical data, and whose outcome is inherently
probabilistic. Given a quantum state $\alpha\ket{0}+\beta\ket{1}$, a
measurement will yield output $0$ with probability $|\alpha|^2$ and
$1$ with probability $|\beta|^2$.  To model this probabilistic effect
in our call-by-value setting, our semantics requires a computational
monad in the sense of {\cite{moggi91notions}}. The coexistence of the
computational monad $T$ and the duplicability comonad $!$ in the same
category is what makes our semantics interesting and novel. It differs
from the work of \cite{benton96linear}, who considered a monad and a
comonad one two different categories, arising from a single
adjunction.

The computational aspects of linear logic have been extensively
explored by many authors, including \cite{bierman93intuitionistic,benton92linear,benton93term,abramsky93computational,wadler92substitute}. However, these works contain explicit
lambda terms to witness the structural rules of linear logic, for
example, $x:{!A}\entail{\rm derelict}(x):A$. By contrast, in our
language, structural rules are implicit at the term level, so that
$!A$ is regarded as a subtype of $A$ and one writes $x:{!A}\entail
x:A$. As we have shown in {\cite{selinger05lambda}}, linearity
information can automatically be inferred by the type checker.  This
allows the programmer to program as in a non-linear language. 

This use of subtyping is the main technical complication in our proof
of well-definedness of the semantics. This is because one has to show
that the denotation is independent of the choice of a potentially
large number of possible derivations of a given typing judgment. We
are forced to introduce a Church-style typing system, and to prove
that the semantics finally does not depend on the additional type
annotations.

Another technical choice we made in our language concerns the relation
between the exponential $\bang{}$ and the pairing operation. Linear
logic only requires $\bang{A}\tensor\bang{B} \entail \bangp{A\tensor
 B}$ and not the opposite implication. However, in our programming
language setting, we find it natural to identify a classical pair of
values with a pair of classical values, and therefore we will have an
isomorphism $\bang{A}\tensor\bang{B} \cong \bangp{A\tensor B}$.

The plan of the paper is the following. First, we describe the lambda
calculus and equational axioms we wish to consider. Then, we develop a
categorical model, called linear category for duplication, which is
inspired by \cite{bierman93intuitionistic} and \cite{moggi91notions}.
We then show that the language is an internal language for the
category, thus obtaining soundness and completeness.

%+++++++++++++++++++++++++++++++++++++++++++++++++++++++++++++++++++++
%++++++++++++++++ SECTION ++++++++++++++++++++++++++++++++++++++++++++
%+++++++++++++++++++++++++++++++++++++++++++++++++++++++++++++++++++++
\section{The language}
\label{sec:language}

We will describe a linear typed lambda calculus with higher-order
functions and pairs. The language is designed to manipulate both
classical data, which is duplicable, and quantum data, which is
non-duplicable. For simplicity, we assume the language is strictly
linear, and not affine linear as in {\cite{selinger05lambda}}. This
means duplicable values are both copyable and discardable, whereas
non-duplicable values must be used once, and only once.
\vspace{-2ex}

%---------------------------------------------------------------------
%---------------- SUBSECTION -----------------------------------------
%---------------------------------------------------------------------
\subsection{The type system}

The set of types is given as follows: ${\it Type}\ A,B ::=  \alpha\bor (A\loli B) 
      \bor (A\tensor B)\bor\putype \bor \bang{A}$.

Here $\alpha$ ranges over type constants. While the remainder of this
paper does not depend on the choice of type constants, 
in our main application~\cite{selinger05lambda}
this is
intended to include a type $\qbit$ of quantum bits, and a type $\bit$
of classical bits. $A\loli B$ stands for
functions from $A$ to $B$, $A\tensor B$ for pairs, $\putype$ for the
unit type, and $\bang{A}$ for duplicable objects of types $A$. We
denote $\bang{\bang{\cdots\bang{A}}}$ with $n$ $\bang{}$'s 
by $\nbang{n}{A}$.

The intuitive definition of $\bang{A}$ is the key to the spirit in
which we want the language to be understood: The $\bang{}$ on
$\bang{A}$ is understood as specifying a property, rather than
additional structure, on the elements of $A$. Therefore, we will have
${!A}\cong{!!A}$. Whether or not a given value of type $A$ is also of
type $!A$ should be something that is inferred, rather than specified
in the code.

Since a term of type $\bang{A}$ can always be seen as a term of type
$A$, we equip the type system with a subtyping relation as follows:
Provided that $(m=0)\vee(n\geq 1)$,
\vspace{-2ex}
\[
\begin{array}{c}
\infer[({\it ax}),]{\nbang{n}{\alpha}\subtype\nbang{m}{\alpha}}{}
\qquad
\infer[(\putype),]{\nbang{n}{\putype}\subtype\nbang{m}{
\putype}}{}
\\
\infer[(\loli),]{
  \nbang{n}{(A'\loli B)}\subtype\nbang{m}{(A\loli B')}}{
  A\subtype A' & B\subtype B'}
\qquad
\infer[(\tensor).]{\
  \nbang{n}{(A\tensor B)}\subtype\nbang{m}{(A'\tensor B')}.}{
  A\subtype A' & B\subtype B'}
\end{array}
\]
This relation encapsulates the main properties terms should satisfy
with respect to duplicability.

%---------------------------------------------------------------------
%---------------- SUBSECTION -----------------------------------------
%---------------------------------------------------------------------
\subsection{Terms}

The language consists of the following typed
terms, divided into \define{values} on the one hand, and general terms, or
\define{computations}, on the other. Both share a subset of the values, the
\define{core values}.
\[
\setlength{\arraycolsep}{0.2ex}
\begin{array}{lll}
  {\it CoreValue}\ U &::= & x^A\bor c^A\bor \puterm^n \bor
      \lambda^n x^A.M, \\[2ex]
  {\it Value}\ V,W &::= & U\bor \prodterm{V,W} \bor 
      \lettermx{x^A}{V}{W}\bor
      \nletprodterm{x^A,y^B}{n}{V}{W}\bor\\&&
      \letunitterm{V}{W},
      \\[2ex]
  {\it Term}\ M,N &::= & U
      \bor\prodterm{M,N}\bor (MN) \bor
      \nletprodterm{x^A,y^B}{n}{M}{N}\bor
      \letunitterm{M}{N},
\end{array}
\]
where $n$ is an integer, $c$ ranges over a set of constant terms, $x$
over a set of term variables and $\alpha$ over a set of constant
types. We abbreviate $(\lambda^0 x^A.M)N$ by $\lettermx{x^A}{N}{M}$,
$\lambda^n x^{\nbang{m}{\putype}}.
\letunitterm{x^{\putype}}{M}$ by $\lambda^n\puterm^m.M$ and
we omit numerical indexes when they are null.

Note that the above terms carry Church-style type annotations, as well as integer superscripts; we call these terms \define{indexed terms}. We also define a notion
of \define{untyped terms} as terms with no index:
\[
\setlength{\arraycolsep}{0.5ex}
\begin{array}{lcl}
  {\it PureTerm}\ M,N&::=& x\bor c\bor \puterm \bor
      \lambda x.M \bor (MN) \bor \prodterm{M,N} \bor\\
      && \letprodterm{x,y}{M}{N}\bor \letunitterm{M}{N}.
\end{array}
\]
The erasure operation $\Erase:{\it Term}\to{\it PureTerm}$ is defined
as the operation of removing the types and integers attached to a
given indexed term.
If $M=\Erase(\bar{M})$, we say that $\bar{M}$ is an \define{indexation
of $M$}

Finally, we define an $\alpha$-equivalence on terms, denoted by
$\eqalpha$, in the usual way (see for 
example \cite{barendregt84lambda}).

%---------------------------------------------------------------------
%---------------- SUBSECTION -----------------------------------------
%---------------------------------------------------------------------
\subsection{Duplicable pairs and pairs of duplicable elements}

Before we formally present the type system, let us informally motivate
our choice of typing rules. One non-obvious choice we had to make is
for the interaction of pairs and duplicability.  Unlike previous works
with comonads \cite{bierman93intuitionistic,benton93term}, we want to
think of the type $\bangp{A\tensor B}$ as a type of pairs of elements
of type $A$ and $B$: we want to use the same operation to access the
components as one would use for a pair of type $A\tensor B$, without
having to use a dereliction operation.

This immediately raises a concern: consider a pair of elements
$\prodterm{x,y}$ of type $\bangp{A\tensor B}$. Are $x$ and $y$
duplicable? In the usual linear logic interpretation, they are not. Having a infinite supply of pair of shoes does not mean
one has an infinite supply of right shoes: we cannot discard the left
shoes. On the other hand, in our interpretation of ``classical'' data as residing in ``classical'' memory and therefore being duplicable, if the string
$\prodterm{x,y}$ is duplicable, then so should be the elements $x$ and $y$.
In other words, we want the duplication to ``permeate'' the pairing.

The choice of such a ``permeable'' pairing is more
or less forced on us by our desire to have no explicit term syntax for
structural rules. Consider the following untyped terms, 
which can be typed if $t$ is of type $\bangp{A\tensor
\bangp{B\tensor C}}$:
\begin{gather}
\letprodterm{x,u}{t}{
  \letprodterm{y,z}{u}{\prodterm{\prodterm{z,y},x}}
},\label{eq:permeable1}\\
\letprodterm{x,u}{t}{
  \prodterm{\letprodterm{y,z}{u}{\prodterm{z,y}},x}
}.\label{eq:permeable2}
\end{gather}
First, we expect these two terms to be axiomatically
equal. Term~\eqref{eq:permeable2} should be of type
$\bangp{\bangp{C\tensor B}\tensor A}$, regardless of the permeability
of the pairing: if $\prodterm{y,z}$ is duplicable, so should be
$\prodterm{z,y}$. Now, consider the term~\eqref{eq:permeable1} with a
non-permeable pairing. In the naive type system, $u$ ends up being of type $B\tensor C$, and the variables $y$ and $z$ in the final
recombination end up being respectively of type $B$ and
$C$. It is not possible to make $\prodterm{z,y}$ of the duplicable type
$\bangp{C\tensor B}$. 

We therefore choose a permeable pairing, which will be reflected,
albeit subtly, in the typing rule $(\otimes.I)$ and $(\otimes.E)$ in
the following section.

%---------------------------------------------------------------------
%---------------- SUBSECTION -----------------------------------------
%---------------------------------------------------------------------
\subsection{Typing judgments}
\label{sec:typjudg}

A typing judgment
is a tuple $\Delta\entail M:A$, where $M$ is an indexed term, $A$ is a
type, and $\Delta$ is a typing context.
To each constant term $c$ we assign a type $\bang{A_c}$.
A \define{valid typing judgment} is a typing judgment that can be
derived from the typing rules in Table~\ref{table:typrules}.
We use the notation $\bang{\Delta}$ for a context where all variables
have a type of the form $\bang{A}$. Finally, when we write a context
$\Gamma,\Delta$, we assume the contexts $\Gamma$ and $\Delta$ to be
disjoint.

%.......... BEGIN typrules ...........................................
\begin{table*}[t]
\def\mynl{\\\\[-1ex]}
\resizebox*{4.2in}{!}{
\begin{minipage}{\textwidth}
\begin{tabular}{c}
\begin{tabular}{c c c}
$$
\infer[({\it ax}_1)]{\bang\Delta, \typ{x}{A} 
  \entail x^B:B}{A \subtype B}
$$ & $$
\infer[({\it ax}_2)]{\bang\Delta \entail c^B:B}{A_c \subtype B}
$$ & $$
\infer[({\it app})]
      {\Gamma_1, \Gamma_2, \bang{\Delta} \entail M N : B}{
        \Gamma_1, \bang{\Delta} \entail M : A \loli B &
        \Gamma_2, \bang{\Delta} \entail N : A
      }
$$
\end{tabular}
\mynl
\begin{tabular}{c c c}
$$
\infer[(\lambda_1)]
      { \Delta \entail \lambda^0 x^A.M : A \loli B}
      { \Delta, \typ{x}{A}\entail M : B}
$$ & $$
\infer[(\lambda_2)]
      { \bang{\Delta} \entail \lambda^{n+1} x^A.M : 
        \nbang{n+1}{(A \loli B)}}
      { \bang{\Delta}, \typ{x}{A} \entail M : B}
$$ & $$
\infer[(\putype.I)]{
  \bang\Delta \entail \puterm^n : \nbang{n}{\putype}
}{}
$$
\end{tabular}
\mynl
\begin{tabular}{c c}
$$
\infer[(\otimes.I)]{
  \bang{\Delta},\Gamma_1,\Gamma_2
  \entail
  \langle M_1,M_2\rangle^n : \nbang{n}{({A_1}\otimes{A_2})}
}{
  \bang{\Delta},\Gamma_1 \entail M_1 : \nbang{n}{A_1} &
  \bang{\Delta},\Gamma_2 \entail M_2 : \nbang{n}{A_2}
}
$$
&
$$
\infer[(\putype.E)]{
  \bang\Delta,\Gamma_1,\Gamma_2\entail
  \letunitterm{M}{N}:A
}{
  \bang\Delta,\Gamma_1\entail M:\putype
  &
  \bang\Delta,\Gamma_2\entail N:A
}
$$
\end{tabular}
\mynl
$$
\infer[(\otimes.E)]{
  \bang\Delta, \Gamma_1, \Gamma_2 
  \entail
  \nletprodterm{x_1^{A_1},x_2^{A_2}}{n}{M}{N} :A
}{
  \bang{\Delta},
  \Gamma_1\entail M:\nbangp{n}{A_1\otimes A_2}
  &
  \bang{\Delta},
  \Gamma_2,x_1:\nbang{n}{A_1},x_2{:}\nbang{n}{A_2}\entail N:A
}
$$
\end{tabular}
\end{minipage}}
\tablenl
\mycaption{Typing rules}
\label{table:typrules}
\end{table*}
%.......... END typrules .............................................

The following lemmas are proved by
structural induction on terms or type derivations as appropriate.

\begin{lemma}\label{lem:duplicvalue}
  If $V$ is a value such that $\Delta\entail V:\bang{A}$ is a valid
  typing judgment, then $\Delta=\bang{\Delta'}$ for some context
  $\Delta'$.
  \qed
\end{lemma}

\begin{lemma}\label{lem:termvar}
  Consider the following valid typing judgment: $\Delta,x:A\entail
  M:B$. Then for every free instance $x^{A'}$ in $M$, $A\subtype A'$.
  \qed
\end{lemma}

\begin{definition}\rm
  In a typing judgment $\Delta\entail M:A$, a term variable
  $x\in|\Delta|$ is called \define{dummy} if $x\not\in\FV(M)$.
\end{definition}

\begin{lemma}\label{lem:dummydup}
  Any dummy variable $x$ in $\Delta\entail M:B$ satisfies
  $\Delta(x)=\bang{A}$,
  for some $A$. Conversely, if $\Delta\entail
  M:B$ is valid and if $x\not\in\FV(M)$, then for all types $A$ the
  typing judgment $\Delta,x:\bang{A}\entail M:B$ is valid.
  \qed
\end{lemma}

Typing derivations are not unique {\it per se}. However for a given
valid typing judgement $\Delta\entail M:A$ two typing derivations will
only differ with respect to the placement of \define{dummy variables},
namely the unused variables in context.

%---------------------------------------------------------------------
%---------------- SUBSECTION -----------------------------------------
%---------------------------------------------------------------------
\subsection{Type casting and substitution Lemma}

\begin{lemma}
\label{lem:typecastvalue}
  Suppose $\Delta\entail M:A$ is a valid typing judgment, and suppose
  $\Delta'\subtype\Delta$ and $A\subtype A'$. Then there exists a
  canonical valid typing judgment $\Delta'\entail M':A'$ such that
  $\Erase(M)=\Erase(M')$. Moreover, if $M$ is a value, so is $M'$.
\end{lemma}

\begin{proof}
  By induction on $M$. \qed
\end{proof}

We will denote this $M'$ with $\s{\Delta'\subtype\Delta\entail
M:A\subtype A'}$. If $\Delta'=\Delta$ or $A'=A$, we omit
them for clarity.

\begin{definition}\label{def:subst}\rm
  Given two valid typing judgments $\bang\Delta,\Gamma_1\entail
  V:A$ and $\bang\Delta,\Gamma_2,x:A\entail M:B$ where $V$ is a value,
  we define the \define{substitution} (with capture avoiding)
  $\bang\Delta,\Gamma_1,\Gamma_2\entail M[V/x]:B$ as follows:
  we replace each free instance $x^{A'}$ (where $A\subtype A'$
  from Lemma~\ref{lem:termvar}) in $M$ by
  $\s{\Delta\entail V:A\subtype A'}$.
\end{definition}

\begin{lemma}[Substitution Lemma]\label{lem:valuesubstisvalue}
  In Definition~\ref{def:subst},
  $\bang\Delta,\Gamma_1,\Gamma_2\entail M[V/x]:B$ is well-typed. Also,
  if $M$ is a value, so is $M[V/x]$.
\end{lemma}

\begin{proof}
  Proof by structural induction on $M$, using
  Lemmas~\ref{lem:termvar} and~\ref{lem:typecastvalue}.\qed
\end{proof}

%- - - - - - - - - - - - - - - - - - - - - - - - - - - - - - - - - - -
%- - - - - - - -  SUBSECTION  - - - - - - - - - - - - - - - - - - -
%- - - - - - - - - - - - - - - - - - - - - - - - - - - - - - - - - - -
\subsection{Axiomatic equivalence}

We define an equivalence relation on (indexed) typing judgments.  We
write $\Delta\entail M\eqax M':A$, or simply $M\eqax M'$, to indicate
that $\Delta\entail M:A$ and $\Delta\entail M':A$ are equivalent.
Axiomatic equivalent is defined as the reflexive, symmetric,
transitive, and congruence closure of the rules from
Tables~\ref{table:eqaxrules}, so
long as both sides of the equivalences are well-typed.
The symbol ``$-$'' is a place holder
for $x$, $\puterm$, or $\prodterm{x,y}$.

%.....................................................................
\begin{table*}[t]
\def\mynl{\\\\[-1ex]}
\resizebox*{4in}{!}{
\begin{minipage}{\textwidth}
\[
\begin{array}{rlll}
  (\axblamb)\quad&
  \Delta\entail \lettermx{x}{V}{M} &\eqax M[V/x] &: A
  \\
  (\axbtens)\quad&
  \Delta\entail \nletprodterm{x,y}{n}{\prodterm{V,W}^n}{M}
  &\eqax M[V/x,W/y] &:A
  \\
  (\axbpu)\quad&
  \Delta\entail \letunitterm{\puterm}{M}
  &\eqax M &:A
  \\
  (\axelamb)\quad&
  \Delta\entail \lambda^n x^A.\s{V:\nbangp{n}{A\loli B}\subtype A\loli
  B}x^A&\eqax V&:\nbangp{n}{A\loli B}.
  \\
  (\axblambA)\quad&
  \Delta\entail \lettermx{x^A}{N}{x^A}&\eqax N&:A.
  \\
  (\axetens)\quad&
  \Delta\entail \nletprodterm{x^A,y^B}{n}{N}{
    \prodterm{x^{\nbang{n}A},y^{\nbang{n}B}}^n
  }
  &\eqax N&:\nbangp{n}{A\tensor B}.
  \\
  (\axepu)\quad&
  \Delta\entail \letunitterm{N}{\puterm^n}
  &\eqax N&:\nbang{n}\putype.
\end{array}
\]
\[
\begin{array}{rlll}
  (\axoletA)\quad&
  \Delta\entail\letterm~{-_1}=(\lettermx{-_2}{M}{N})~\interm P
  \hspace{-4ex}&\eqax
  \letterm~-_2=M\interm\lettermx{-_1}{N}{P}&:A
  \\
  (\axoletB)\quad&
  \Delta\entail\letterm~-_1=V\interm\lettermx{-_2}{W}{M}
  &\eqax
  \letterm~-_2=W\interm\lettermx{-_1}{V}{M}&:A
  \\
  (\axoletappB)\quad&
  \Delta\entail \letterm~x^{A\loli B}=M\interm\lettermx{y^A}{N}{xy} &\eqax
  MN&:B
  \\
  (\axoletlamb)\quad&
  \Delta\entail\lettermx{x^D}{V}{\lambda^n y^A.M}&\eqax
  \lambda^n y^A.\lettermx{x^D}{V}{M}&:\nbangp{n}{A\loli B}
  \\
  (\axolettensC)\quad&
  \Delta\entail \letterm~x^A=M\interm\lettermx{y^B}{N}{\prodterm{x^A,y^B}^n}
  \hspace{-3ex}&\eqax \prodterm{M,N}^n&:\nbangp{n}{A\tensor B}
\end{array}
\]
\[
\begin{array}{rl}
 (\axtapp)\quad&\s{M:\nbangp{n}{A\loli D}\subtype 
    B\loli D'}\s{N:C\subtype B}\\
  &\qquad\eqax
  \s{\s{M:\nbangp{n}{A\loli D}\subtype A\loli D}
    \s{N:C\subtype A}:D\subtype D'}
  \\
  (\axtpair)\quad&\nletprodterm{x^{A'},y^{B'}}{n'}{\s{M : 
      \nbangp{n}{A\tensor B}\subtype \nbangp{n'}{A'\tensor B'}}}{N}\\
  &\qquad\eqax \nletprodterm{x^A,y^B}{n}{M}{
    \s{\Delta,x:\nbang{n}A,y:\nbang{n}B\subtype \Delta,x:\nbang{n'}A',
      y:\nbang{n'}B'\entail N}
  }
  \\
  (\axtletx)\quad&\lettermx{x^{A'}}{\s{M:A\subtype A'}}{N}
  \eqax \lettermx{x^A}{M}{\s{\Delta,x:A\subtype \Delta,x:A'\entail N}}
  \\
  (\axtletu)\quad&\letunitterm{\s{M:\nbang{m}{\putype}\subtype
      \nbang{n}{\putype}}}{N}
  \eqax \letunitterm{M}{N}
\end{array}
\]
\end{minipage}}
\tablenl
\mycaption{Axiomatic equivalence axioms}
\label{table:eqaxrules}
\end{table*}
%.....................................................................

\begin{lemma}
  The equivalences of
  Table~\ref{table:eqaxderived} are 
  derivable. \qed
\end{lemma}

%.....................................................................
\begin{table*}[t]
\resizebox{4.2in}{!}{
\begin{minipage}{\textwidth}
\begin{align*}
  (\axalet)\quad &\Delta,x:A\entail \lettermx{y^A}{x^A}{M}:B &&\eqax
  \Delta,y:A\entail M&&:B
  \\[-1ex]
  (\axoletlambB)\quad&
  \bang\Delta\entail\lettermx{x^{\bang{C}}}{V}{\lambda y.M}
  &&\eqax
  \lambda^{n+1} y.\lettermx{x^{\bang{C}}}{V}{M}
  &&:\nbangp{n+1}{A\loli B}
  \\[-1ex]
  (\axolettensB)\quad&
  \Delta\entail\prodterm{V,\lettermx{-}{M}{N}} &&\eqax
  \lettermx{-}{M}{\prodterm{V,N}}&&:\nbangp{n}{A\tensor B}
  \\[-1ex]
  (\axolettensA)\quad&
  \Delta\entail\prodterm{\lettermx{-}{M}{N},V} &&\eqax
  \lettermx{-}{M}{\prodterm{N,V}}&&:\nbangp{n}{A\tensor B}
  \\[-1ex]
  (\axoletapp)\quad&
  \Delta\entail V(\lettermx{-}{M}{N}) &&\eqax
  \lettermx{-}{M}{VN}&&:B
  \\[-1ex]
  (\axoletappA)\quad&
  \Delta\entail (\lettermx{-}{M}{N})V &&\eqax
  \lettermx{-}{M}{NV}&&:B  
\end{align*}
\end{minipage}
}
\tablenl
\mycaption{Axiomatic equivalence: derived rules}
\label{table:eqaxderived}
\end{table*}
%.....................................................................

The following result stipulates that all the indexations of a given
erasure live in the same axiomatic class. In other words, the
axiomatic equivalence class of a term is independent of its
indexation.

\begin{theorem}\label{the:eraseimpliesax}
  If $\Erase(M)=\Erase(M')$ and if $\Delta\entail M,M':A$ are valid
  typing judgments, then $M\eqax M'$.
\end{theorem}

\begin{proof}[Sketch] 
  The actual proof is long and technical, and is omitted here for
  space reasons. We proceed by first defining a special subset of
  terms, called {\em neutral} terms, for which the Theorem is obvious.
  We then prove that every term is axiomatically equivalent to a
  neutral term via a series of rewrite systems.\qed
\end{proof}

%+++++++++++++++++++++++++++++++++++++++++++++++++++++++++++++++++++++
%++++++++++++++++ SECTION ++++++++++++++++++++++++++++++++++++++++++++
%+++++++++++++++++++++++++++++++++++++++++++++++++++++++++++++++++++++
\section{Linear category for duplication}
\label{sec:lincat}

As it was advertised, the structure of the categorical semantics will
closely follow the one proposed by Bierman~\cite{bierman93intuitionistic}, but
with the added twist of a computational monad {\`a} la
Moggi~\cite{moggi91notions}.
Indeed, since one has tensor product and a tensor unit, one can
expect the categorical model to be symmetric monoidal. Since
one can construct candidate maps for building a comonad, a comonoid
structure for each $\bang{A}$ and coherence maps for the comonad, we
have a linear category.
Finally, the computational aspect will be taken care by Moggi's
computational monad.

%---------------------------------------------------------------------
%---------------- SUBSECTION -----------------------------------------
%---------------------------------------------------------------------
\subsection{Linear exponential comonads}

In his Ph.D. thesis, Bierman~\cite{bierman93intuitionistic} gives the
definition of a \define{linear category}. We prefer here the
terminology given in \cite{schalk04what}, and use the concept of
\define{linear exponential comonad}.

\begin{definition}\rm
\label{def:lincate}
  Let $(\cat{C},\tensor,\putype)$ be a symmetric monoidal
  category~\cite{maclane98categories},
  where
  $\alpha_{A,B,C}: A\tensor(B\tensor C) \to (A\tensor B)\tensor C$,
  $\lambda_A     : \tensunit\tensor A\to A$,
  $\rho_A        : A\tensor\tensunit \to A$ and
  $\sigma_{A,B}  : A\tensor B\to B\tensor A$
  are the usual associativity, left unit, right unit and symmetry
  morphisms. Let
  $(L,\comult,\counit,\cohmap{L},\cohmap{L})$ be a monoidal
  comonad~\cite{bierman93intuitionistic},
  where $\counit_A:LA\to A$, 
    $\comult_A:LA\to LLA$,
    $\cohmap{L}_{A,B} :LA\tensor LB \to L(A\tensor B)$ and
    $\cohmap{L}_\putype:\tensunit\to L\tensunit$.
  We say that $L$ is a \define{linear exponential
  comonad}~\cite{schalk04what} provided that
  \begin{enumerate}
  \item each object in $\cat{C}$ of the form $LA$ is equipped with a
    commutative comonoid $(LA,\comonoidmult_A,\comonoidunit_A)$,
    where
      $\comonoidmult_A:LA\to LA\tensor LA$ and
      $\comonoidunit_A:LA\to \putype$;
  \item
    $\comonoidmult_A$ and
    $\comonoidunit_A$ are monoidal natural
    transformations;
  \item $\comonoidmult_A:(LA,\comult_A)\to(LA\tensor LA,
    (\comult_A\tensor\comult_A);d_A)$ and
    $\comonoidunit_A:(LA,\comult_A)\to 
    (\putype,\cohmap{L}_{\putype})$ are
    $L$-coalgebra morphisms;
  \item Every map $\comult_A$ is a comonoid morphism
    $(LA,\comonoidunit_A,\comonoidmult_A) \to
    (L^2A,\comonoidunit_{LA},
    \comonoidmult_{LA})$.
  \end{enumerate}
  The equations for $2$--$4$ are to be found in
  Table~\ref{table:eqlinexpcom}.
\end{definition}

%.....................................................................
\begin{table*}[t]
  \resizebox{4.2in}{!}{
  \begin{minipage}{\textwidth}
    \[
    \def\t{\tensor}\def\u{\putype}
    \begin{array}{c}
    \begin{array}{c}
    \xymatrix@C=10ex@R=6ex{
        LA\t LB
        \ar[dd]_{\cohmap{L}_{A,B}}
        \ar[r]^<(0.2){\comonoidmult_A\t\comonoidmult_B}
      & (LA\t LA)\t(LB\t LB)
        \ar[d]^{\swapmap}
        \\
      & (LA\t LB)\t(LA\t LB)
        \ar[d]^{\cohmap{L}_{A,B}\t\cohmap{L}_{A,B}}
        \\
        L(A\t B) 
        \ar[r]_<(0.2){\comonoidmult_{(A\t B)}}
      & L(A\t B)\t L(A\t B)
    }
    \end{array}
    \ 
    \begin{array}{c}
    \xymatrix@C=2ex@R=4ex{
         \u
         \ar[r]^<(0.3){\lambda^{-1}_{\u}}
         \ar[d]_{\cohmap{L}_{\u}}
       & \u\t\u 
         \ar[d]^{\cohmap{L}_{\u}\t\cohmap{L}_{\u}}
         \\
         L\u
         \ar[r]_<(0.3){\comonoidmult_{\u}}
       & L\u\t L\u
    }
    \ 
    \xymatrix@C=5ex@R=4ex{
        LA\t LB
        \ar[r]^<(0.3){\comonoidunit_A\t\comonoidunit_B}
        \ar[d]_{\cohmap{L}_{A,B}}
      & \u\t\u
        \ar[d]^{\lambda_{\u}}
        \\
        L(A\t B)
        \ar[r]_<(0.3){\comonoidunit_{A\t B}}
      & \u
    }
    \\
    \xymatrix@C=8ex@R=4ex{
      \u \ar[rr]^{\id}\ar[rd]_{\cohmap{L}_{\u}} && \u;\\
      &L\u\ar[ru]_{\comonoidunit_{\u}} &
    }
    \end{array}
    \\
    \textrm{$\comonoidmult_A$ and $\comonoidunit_A$
      are monoidal natural transformations}.
    \end{array}
    \]
    \[
    \def\t{\tensor}\def\u{\putype}
    \begin{array}{cc}
      \begin{array}{c}
    \xymatrix@R=4ex{
        LA \ar[r]^{\comonoidmult_A}
        \ar[dd]_{\comult_A}
      & LA\t LA
        \ar[d]^{\comult_A\t\comult_A}
        \\
      & L^2A\t L^2A
        \ar[d]^{\cohmap{L}_{LA,LA}}
        \\
        L^2A
        \ar[r]_{L\comonoidmult_A}
      & L(LA\t LA),
    }
    \ 
    \xymatrix@R=12ex{
        LA \ar[r]^{\comonoidunit_A}
        \ar[d]_{\comult_A}
      & \u
        \ar[d]^{\cohmap{L}_{\u}}
        \\
        L^2A
        \ar[r]_{L\comonoidunit_A}
      & L\u;
    }
      \end{array}
    &
    \begin{array}{c}
    \xymatrix@R=5ex{
        LA \ar[r]^{\comult_A}
        \ar[d]_{\comonoidmult_A}
      & L^2A
        \ar[d]_{\comonoidmult_{LA}},
        \\
        LA\t LA \ar[r]_{\comult_A\t\comult_A}
      & L^2A\t L^2A
    }
    \xymatrix@C=2ex@R=5ex{
       LA \ar[rr]^{\comult_A}
       \ar[dr]_{\comonoidunit_A}
     &&L^2A.
       \ar[dl]^{\comonoidunit_{LA}}
       \\
       &\u&
    }
    \end{array}
    \\
    \textrm{$\comonoidmult_A$ and $\comonoidunit_A$ are $L$-coalgebra maps.}
    &
    \textrm{$\comult_A$ us a comonoid morphism.}
    \end{array}
    \]
  \end{minipage}
  }
\tablenl
\mycaption{Equations for a linear exponential comonad}
\label{table:eqlinexpcom}
\end{table*}
%.....................................................................

%---------------------------------------------------------------------
%---------------- SUBSECTION -----------------------------------------
%---------------------------------------------------------------------
\subsection{Strong monad and $T$-exponentials.}

To capture the computational effect of the probabilistic measurement,
we use the notion of \define{strong monad}, as in
\cite{moggi91notions}. Recall that
  a \define{monad} over a category $\cat{C}$ is a triple
  $(T,\eta,\mu)$ where $T:\cat{C}\to\cat{C}$ is a functor,
  $\eta:\id\nt T$ and $\mu:T^2\nt T$ are natural transformations and
  such that $T\mult_A;\mult_A=\mult_{TA};\mult_A$ and $\unit_{TA};\mult_A=\id_{TA}=T\unit_A;\mult_A$.
Given a map $f:A\to TB$, we define the map $f^*:TA\to TB$ by
$Tf;\mult_B$.

\begin{definition}\label{def:strongmonad}\rm
  A \define{strong monad} over a monoidal category $\cat{C}$
  is a monad $(T,\eta,\mu)$ together with a natural transformation
  $t_{A,B}:A\tensor TB\to T(A\tensor B)$, called the \define{tensorial
  strength}, subject to a number of coherence conditions. 
\end{definition}

\begin{remark}
  If the category $\cat{C}$ is symmetric, the tensorial strength
  $t$ induces two natural transformations
  $TA\tensor TB\to T(A\tensor B)$, namely
  \[\begin{array}{ll}
    \Strength_1:TA\tensor TB\xrightarrow{\sigma_{TA,TB}}
    TB\tensor TA\xrightarrow{t_{TB,A}}
    T(TB\tensor A)
    \xrightarrow{(\sigma_{TB,A};t_{A,B})^*}
    T(A\tensor B),
    \\
    \Strength_2:TA\tensor TB\xrightarrow{t_{TA,B}}
    T(TA\tensor B)
    \xrightarrow{(\sigma_{TA,B};t_{B,A})^*}
    T(B\tensor A) \xrightarrow{T\sigma_{B,A}}
    T(A\tensor B).
  \end{array}\]
  Note that $\Strength_1$ and $\Strength_2$ might not
  be equal: the map $\Strength_1$
  ``evaluates'' the first variable and then the second one. The map
  $\Strength_2$ does the opposite.  The strength is called
  \define{commutative} if $\Strength_1=\Strength_2$.
\end{remark}

\begin{lemma}\label{lem:strongmonadmonoidal}
  If $(T,\unit,\mult,t)$ is a strong monad on a symmetric monoidal
  category $\cat{C}$, then $(T,\unit,\mult,\Psi_1)$ and 
  $(T,\unit,\mult,\Psi_2)$ are monoidal monad.\qed
\end{lemma}

\begin{definition}\label{def:texponential}\rm
  A symmetric monoidal category $(\cat{C},\tensor,\tensunit)$ together
  with a strong monad $(T,\eta,\mu)$ is said to have
  \define{$T$-exponentials}~\cite{moggi91notions},
  or \define{Kleisli exponentials},
  if it is equipped
  with a bifunctor $\loli:\cat{C}^{\it op}\times\cat{C}\to\cat{C}$,
  and a natural isomorphism
  \[
  \Phi:\cat{C}(A, B\loli C) \xrightarrow{\quad\cong\quad}
  \cat{C}(A\tensor B, TC).
  \]
\end{definition}

\begin{lemma}
  The map $\Phi$ induces a natural transformation
  $\appnt_{A,B} : (A\loli B)\tensor A\to TB$
  defined by $\Phi(\id_{A\loli B})$. \qed
\end{lemma}

%---------------------------------------------------------------------
%---------------- SUBSECTION -----------------------------------------
%---------------------------------------------------------------------
\subsection{Idempotent, strong monoidal comonad}

A comonad $(L,\counit,\comult)$ on some category is said to be
\define{idempotent} if $\comult:L\nt LL$ is an isomorphism.
A monoidal comonad $(L,\comult,\counit,\cohmap{L},\cohmap{L})$ is
\define{strong monoidal} if $\cohmap{L}_{\putype}$ and $\cohmap{L}_{A,B}$ are
isomorphisms.

\begin{definition}\label{def:idempcanon}\rm
  Given a monoidal category $(\cat{C},\tensor,\tensunit)$ with
   an idempotent, strong monoidal comonad $(L,\counit,\comult)$,
   a bifunctor
    $\loli:\cat{C}^{\it op}\times\cat{C}\to \cat{C}$,
  we define a \define{canonical arrow for $\cat{C}$ with respect to
  duplication} by induction:
  For all objects $A$, the arrows $\id_A$, $\counit_A$,
    $\comult_A$, $\cohmap{L}_{\tensunit}$ and $\cohmap{L}_{A,B}$
    are canonical.
  All \define{expansions of canonical arrows with respect to
    duplication} are also canonical. An expansion of an arrow
    $f:A\to B$ is defined to be either $f$ or any of
    $Lg$,$ X\tensor g$, $g\tensor X$, $X\loli g$, $g\loli X$,
    where $g$ is an expansion of $f$ and $X$ ranges over the objects
    of the category.
  Finally, compositions of canonical arrows are also canonical.
\end{definition}

\begin{theorem}[Coherence for idempotent comonads]\label{the:ciposet}
  Given a category $\cat{C}$ with the structure in
  Definition~\ref{def:idempcanon}, if $f,g:A\to B$ are two canonical
  arrows with respect to duplication, then they are equal. \qed
\end{theorem}

%---------------------------------------------------------------------
%---------------- SUBSECTION -----------------------------------------
%---------------------------------------------------------------------
\subsection{Linear category for duplication}

We now have enough background to define a candidate for the
categorical model of the language we describe in
Section~\ref{sec:language}.

\begin{definition}\rm
  A \define{linear category for duplication} is a
  category $\cat{C}$ with the following structure:
  \begin{itemize}
  \item a symmetric monoidal structure
  $(\tensor,\tensunit,\alpha,\lambda,\rho,\sigma)$;
  \item an idempotent, strongly monoidal, linear exponential comonad
  $(L,\comult,\counit,\cohmap{L},\cohmap{L},$ 
    $\comonoidunit,\comonoidmult)$;
  \item a strong monad $(T,\mult,\unit)$;
  \item a Kleisli exponential $\loli$.
  \end{itemize}
  The \define{computational linear category} is defined to be
  the Kleisli category $\cat{C}_T$, as defined in~\cite{moggi91notions}.
\end{definition}

\begin{remark}
  A linear category for duplication gives rise to a double adjunction
  \resizebox{!}{0.3in}{
  $\xymatrix{
    \kcat{C}{L}\ar@/^3ex/[rr]^{U^L}
    &\bot& \cat{C}\ar@/^3ex/[ll]^{F^L}
    \ar@/^3ex/[rr]^{U^T}
    &\bot& \kcat{C}{T},\ar@/^3ex/[ll]^{F^T}.
  }$
  }
  Here the left adjunction arises from the co-Kleisli category
  $\kcat{C}{L}$ of the comonad $L$. It is as in the linear-non-linear
  models of {\cite{benton94mixed}}, and $\kcat{C}{L}$ is a category of
  classical (non-quantum) values. The right adjunction arises from the
  Kleisli category $\kcat{C}{T}$ of the computational monad $T$, as in
  \cite{moggi91notions}. Here $\kcat{C}{T}$ is a category of
  (effectful) quantum computations.
\end{remark}

%---------------------------------------------------------------------
%---------------- SUBSECTION -----------------------------------------
%---------------------------------------------------------------------
\subsection{The category $\YAQ$}
\label{sec-yaq}

%.....................................................................
\begin{table*}[t]
\resizebox*{4.8in}{!}{
\begin{minipage}{\textwidth}
\[
\begin{array}{llrll}
  \alpha_{A,B,C} &=& x:A{\tensor}(B{\tensor}C)&\entail
  \letterm\prodterm{y,z}=x\interm
  \letterm\prodterm{t,u}=z\interm
  \prodterm{\prodterm{y,t},u}&:(A{\tensor}B){\tensor} C
  \\
  \lambda_A &=& x:\tensunit\tensor A &\entail
  \letprodterm{y,z}{x}{\letunitterm{y}{z}}&:A
  \\
  \rho_A&=& x:A\tensor\tensunit &\entail
  \letprodterm{y,z}{x}{\letunitterm{z}{y}}&:A
  \\
  \sigma_{A,B}&=&x:A\tensor B&\entail
  \letprodterm{y,z}{x}{\prodterm{z,y}}&:B\tensor A
  \\
  \unit_A &=& x:A&\entail\lambdau.x&:\comptype A
  \\
  \mult_A &=& x:\comptype(\comptype A)&\entail
  \lambdau.(x\puterm)\puterm&: \comptype A
  \\
  \strength_{A,B} &=& z:A\tensor(\comptype B)&\entail
  \letprodterm{x,y}{z}{\lambdau.\prodterm{x,y\puterm}}&:
  \comptype(A\tensor B)
  \\
  \counit_A&=& x:\bang{A}&\entail x^A&:A
  \\
  \comult_A&=& x:\bang{A}&\entail x^{\nbang{2}{A}}&:\nbang{2}{A}
  \\
  \cohmap{\bang{}}_{A,B} &=&
  z:\bang{A}\tensor\bang{B}&\entail
  \letprodterm{x,y}{z}{\prodterm{x,y}}&: \bangp{A\tensor B}
  \\
  \cohmap{\bang{}}_{\putype} &=&
  z:\putype&\entail \letunitterm{z}{\puterm}&:\bang{\putype}
  \\
  \comonoidmult_A&=&x:\bang{A}&\entail
  \prodterm{x,x}&:\bang{A}\tensor\bang{A}
  \\
  \comonoidunit_A&=&x:\bang{A}&\entail
  \puterm&:\putype
\end{array}
\]
\[
\begin{array}{lcrll}
  (x:A\entail V:B)\tensor(y:C\entail W:D)
  &=& z:A\tensor B&\entail \letprodterm{x,y}{z}{\prodterm{V,W}}
  &:C\tensor D
  \\
  (x:A\entail V:B)\loli(y:C\entail W:D) &=& z:B\loli C&\entail
  \lambda x.(\lettermx{y}{zV}{W})&:A\loli D
  \\
  (x:A\entail V:\comptype B)^*&=&
  y:\comptype A&\entail \lambdau.\lettermx{x}{(y\puterm)}{(V\puterm)}
  &:\comptype B
  \\
  \Phi_{A,B,C} \left(x:A\entail V:B\loli C\right) & = &
  t:A\tensor B&\entail \lambda\puterm.\letprodterm{x,y}{t}{Vy}
  &:\putype\loli C
\end{array}
\]
\end{minipage}}
\tablenl
\mycaption{Definitions of maps and operations on maps in $\YAQ$}
\label{table:yaqmap}
\end{table*}
%.....................................................................

\begin{definition}\label{def:yaq}\rm
  We can define a category $\YAQ$ as follows: Objects are types,
  and arrows $A\to B$ are axiomatic classes of valid typing judgments of
  the form $x:A\entail V:B$, where $V$ is a value. We define the
  composition of arrows $x:A\entail V:B$ and $y:B\entail W:C$ to be
  $x:A\entail\lettermx{y}{V}{W}:C$. The identity on $A$ is set to be
  the arrow $x:A\entail x:A$.
\end{definition}

\begin{lemma}\label{lem:yaq}
  The category $\YAQ$ is well-defined.
\end{lemma}

\begin{proof}
  The composition of two arrows yields an arrow axiomatically
  equivalent to a value due to Axiom $(\axblamb)$ and
  Lemma~\ref{lem:valuesubstisvalue}. Composition is associative due to
  Axiom $(\axoletA)$. The arrow $x:A\entail x:A$ is indeed the
  identity on $A$ due to axioms $(\axalet)$ and $(\axblambA)$.
  \qed
\end{proof}

\begin{lemma}\label{lem:promote}
  Given a valid typing judgment $\Delta\entail V:A$ where $V$ is a
  value, there exists a canonical value $V'$ such that
  $\Erase(V')=\Erase(V)$ and such that $\bang\Delta\entail
  V':\bang{A}$. We denote this $V'$ by
  $\s{\bang\Delta\subtype\Delta\entail V:A\overtype A'}$.
\end{lemma}

\begin{proof}
  By induction on $V$.\qed
\end{proof}

\begin{lemma}\label{lem:bangexaq}
  If $\Delta\entail V\eqax W:A$, and if $V'=\s{\bang\Delta\subtype\Delta
  \entail V:A\overtype A'}$ and $W'=\s{\bang\Delta\subtype\Delta
  \entail W:A\overtype A'}$, then $V'\eqax W'$.
\end{lemma}

\begin{proof}
  By induction on $V\eqax W$.\qed
\end{proof}

\begin{theorem}
  If we define $T(A) := \putype\loli A$ and $L(A)=!A$, together with
  the maps and the operations on maps defined in
  Table~\ref{table:yaqmap}, $\YAQ$ is a linear category for
  duplication.
\end{theorem}

\begin{proof}
  The proof is mainly a long list of verifications. It uses
  Theorem~\ref{the:eraseimpliesax}, Lemmas~\ref{lem:yaq},
  \ref{lem:promote} and~\ref{lem:bangexaq}.\qed
\end{proof}

%+++++++++++++++++++++++++++++++++++++++++++++++++++++++++++++++++++++
%++++++++++++++++ SECTION ++++++++++++++++++++++++++++++++++++++++++++
%+++++++++++++++++++++++++++++++++++++++++++++++++++++++++++++++++++++
\section{Denotational semantics}

%---------------------------------------------------------------------
%---------------- SUBSECTION -----------------------------------------
%---------------------------------------------------------------------
\subsection{Interpretation of the language}

The lambda-calculus defined in Section~\ref{sec:language} is thought
as a \define{computational} lambda-calculus. Using Moggi's technique, 
we split the interpretation of
the language into the interpretation of the \define{values} in a
linear category for duplication $\cat{C}$ and the interpretation of
the \define{computations}, i.e. general terms, in its Kleisli
category $\kcat{C}{T}$. Without loss of generality, for notation
purposes, we assume the category to be strictly monoidal.

We define an \define{interpretation of the type system} to be a
map $\Theta$ that assigns to each constant type $\alpha$ an object
$\Theta(\alpha)$.
Each type $A$ is interpreted as an object of
$\cat{C}$:
$\denot{\alpha}_\Theta = \Theta(\alpha)$,
$\denot{\putype}_\Theta = \tensunit$, 
$\denot{\bang{A}}_\Theta = L{\denot{A}_\Theta}$,
$\denot{A\tensor B}_\Theta =
\denot{A}_\Theta\tensor\denot{B}_\Theta$ and
$\denot{A\loli B}_\Theta
= \denot{A}_\Theta\loli\denot{B}_\Theta$.

Given a valid subtyping $A\subtype B$, there exists a canonical arrow
$\denot{A}_{\Theta}\to \denot{B}_{\Theta}$ in $\cat{C}$ with respect
to duplication, as defined in Definition~\ref{def:idempcanon}.
Moreover, this arrow is unique by Theorem~\ref{the:ciposet}.  We
extend the map $\Theta$ to interpret $A\subtype B$ as this unique
arrow and we denote it by $I_{A,B}$.

We use the following straightforward shortcut definitions, where
$A,A',B,B'$ are types and $\Delta$, $\Gamma$ and $\Gamma'$ are typing contexts:
\begin{itemize}
\item $\Split_{\bang\Delta,\Gamma,\Gamma'}:\denot{\bang\Delta}
  \tensor\denot{\Gamma}\tensor\denot{\Gamma'}
  \to \denot{\bang\Delta}\tensor\denot{\Gamma}\tensor
  \denot{\bang\Delta}\tensor\denot{\Gamma'}$.
\item Given $f : \denot{\bang\Delta}\tensor\denot{\Gamma}\to\denot{A}$
  and $g : \denot{\bang\Delta}\tensor\denot{\Gamma'}\to\denot{B}$, we
  define the map
  $f\tensor_{\bang\Delta}g:\denot{\bang\Delta}
  \tensor\denot{\Gamma}\tensor\denot{\Gamma'} \to
  A\tensor B$.
\item Given a natural transformation $n_A:FA\to GA$, if
  $\Delta=\s{x_1:A_1\ldots x_n:A_n}$ we define
  $n_{\Delta}=n_{\denot{A_1}}\tensor\ldots\tensor n_{\denot{A_n}}$.
\end{itemize} 

\begin{definition}\label{def:langdenot}\rm
  The map $\Theta$ is said to be an \define{interpretation of the
  language} if moreover it assigns to each constant term $c:A_c$ an
  arrow $\Theta(c):\tensunit\to \denot{A_c}$ in $\cat{C}$.
  
  Given a linear category for duplication $\cat{C}$, it is possible to
  interpret the typing derivation of a well-typed value as a map in
  $\cat{C}$ and the typing derivation of a valid computation as a map
  in the Kleisli category $\kcat{C}{T}$. We define them inductively.
  \begin{itemize}
  \item
    If $x_1:A_1,\ldots x_n:A_n\entail V:B$ is a value with
    typing derivation $\pi$, its \define{value interpretation}
    $\denot{\pi}_\Theta^v$ is an arrow
    $\denot{A_1}\tensor\ldots\tensor\denot{A_n}\to_{\cat{C}}
    \denot{B}$;
  \item
    if $x_1:A_1,\ldots x_n:A_n\entail M:A$ is a term with typing
    derivation $\pi$, its \define{computational interpretation}
    $\denot{\pi}_\Theta^c$ is an arrow
    $\denot{A_1}\tensor\ldots\tensor\denot{A_n}\to_{\cat{C}}
    T(\denot{B})$.
  \end{itemize}
  Table~\ref{table:modelcorevaluejudg} formulates the definition in
  the simple case where the contexts $\Delta$, $\Gamma_1$ and
  $\Gamma_2$ contain only one variable. One can easily extend this
  to the general setting. 
\end{definition}

%.......... BEGIN modelcorevaluejudg .................................
\begin{table*}[t]
\def\mynl{\\\\[-1ex]}
\resizebox{4.6in}{!}{
\begin{minipage}{6.5in}
Interpretation of core values:
\vspace{-2ex}
\[
\begin{array}{c}
\begin{array}{cc}
\begin{array}{ll}
  \denot{\bang\Delta,x:A\entail x:B}_\Theta^v
&= \denot{\bang\Delta}\tensor\denot{A}
  \xrightarrow{\comonoidunit_{\Delta}\tensor I_{A,B}}\denot{B}
\\
  \denot{\bang\Delta\entail c:B}_\Theta^v
&= \denot{\bang\Delta}
  \xrightarrow{\comonoidunit_{\Delta}}\tensunit
  \xrightarrow{\Theta(c)} \denot{A_c}
  \xrightarrow{I_{A_c,B}}\denot{B}
\\
  \denot{\bang\Delta\entail \puterm:\nbang{n}{\putype}}_\Theta^v
&= \denot{\bang\Delta}
  \xrightarrow{\comonoidunit_{\Delta}}
  \tensunit\xrightarrow{\cohmap{L}_{\tensunit}}
  L\tensunit\xrightarrow{I_{\bang\putype,\nbang{n}{\putype}}}
  L^n\tensunit
\end{array}
&\qquad
\begin{array}{rl}
  \denot{\Delta,x:A\entail M:B}_\Theta^c
&= \denot{\Delta}\tensor\denot{A}
  \xrightarrow{f} T(\denot{B})
\\\hline
  \denot{\Delta\entail\lambda x.M:A\loli B}_\Theta^v
&=\denot{\Delta}
  \xrightarrow{\Phi^{-1}(f)} \denot{A}\loli\denot{B}
\end{array}
\end{array}
\mynl
\begin{array}{c}
  \denot{\bang\Delta,x:A\entail M:B}_\Theta^c
= \denot{\bang\Delta}\tensor\denot{A}
  \xrightarrow{f} T(\denot{B})
\\\hline
  \denot{\bang\Delta\entail\lambda
    x.M:\nbangp{n+1}{A\loli B}}_\Theta^v
= \denot{\bang\Delta}
  \xrightarrow{L(\Phi^{-1} f);I_{\bangp{A\loli B},\nbangp{n+1}{A\loli B}}}
  L^{n+1}(\denot{A}\loli\denot{B})
\end{array}
\end{array}
\vspace{-1.5ex}
\]
\vspace{-1.5ex}
Interpretation of extended values:
\[
\begin{array}{l}
\begin{array}{ll}
  \denot{\bang\Delta,\Gamma_1\entail V:A}_\Theta^v
 = \denot{\bang{\Delta}}{\tensor}\denot{\Gamma_1}
  \xrightarrow{f} \denot{A}
&\qquad
  \denot{\bang\Delta,\Gamma_2,x:A\entail W:B}_\Theta^v
 = \denot{\bang\Delta}{\tensor}\denot{\Gamma_2}{\tensor}\denot{A}
  \xrightarrow{g} \denot{B}
\end{array}
\\
\hline
 \denot{\bang\Delta,\Gamma_2,\Gamma_1\entail 
    \lettermx{x}{V}{W}:B}_\Theta^v
=
\denot{\bang\Delta}{\tensor}\denot{\Gamma_2}{\tensor}\denot{\Gamma_1}
  \xrightarrow{\id{\tensor}_{\bang\Delta}f}
  \denot{\bang\Delta}{\tensor}\denot{\Gamma_2}{\tensor}\denot{A}
  \xrightarrow{g}
  \denot{B}
\end{array}
\]
\[
\begin{array}{l}
\begin{array}{ll}
  \denot{\bang\Delta,\Gamma_1,
  \entail V:\nbangp{n}{A_1\tensor A_2}}_\Theta^v
&= \denot{\bang\Delta}{\tensor}\denot{\Gamma_1}
  \xrightarrow{f} L^n(\denot{A_1}\tensor\denot{A_2})
\\
 \denot{\bang\Delta,\Gamma_2,
   x:\nbang{n}A_1,y:\nbang{n}A_2\entail W:C}_\Theta^v
&= \denot{\bang\Delta}{\tensor}\denot{\Gamma_2}{\tensor}
   L^n\denot{A_1}\tensor L^n\denot{A_2}
  \xrightarrow{g} \denot{C}
\end{array}
\\
\hline
  \denot{\bang\Delta,\Gamma_2,\Gamma_1\entail
  \nletprodterm{x,y}{n}{V}{W}:C}_\Theta^v
 = \denot{\bang\Delta}{\tensor}\denot{\Gamma_2}{\tensor}\denot{\Gamma_1}
 \xrightarrow{\id\tensor_{\bang\Delta} f}
  \denot{\bang\Delta}{\tensor}\denot{\Gamma_2}{\tensor}L^n(\denot{A_1}{\tensor}\denot{A_2})
\\
\qquad\qquad
\xrightarrow{\id\tensor
  \left(\cohmap{L^n}_{\denot{A_1},\denot{A_2}}\right)^{-1}}
  \denot{\bang\Delta}{\tensor}\denot{\Gamma_2}{\tensor}L^n\denot{A_1}
  {\tensor}L^n\denot{A_2}
  \xrightarrow{g}\denot{C}
\end{array}
\]
\[
\begin{array}{l}
\begin{array}{ll}
  \denot{\bang\Delta,\Gamma_2\entail V:\putype}_\Theta^v
 = \denot{\bang\Delta}\tensor\denot{\Gamma_2}
  \xrightarrow{f} \tensunit
&\qquad
  \denot{\bang\Delta,\Gamma_1\entail W:C}_\Theta^v
 = \denot{\bang\Delta}\tensor\denot{\Gamma_1}
  \xrightarrow{g} \denot{C}
\end{array}
\\
\hline
\denot{\bang\Delta,\Gamma_1,\Gamma_2\entail
  \letunitterm{V}{W}:C}_\Theta^v
=
\denot{\bang\Delta}{\tensor}\denot{\Gamma_1}{\tensor}\denot{\Gamma_2}
\xrightarrow{\id\tensor_{\bang\Delta} f}
\denot{\bang\Delta}\tensor\denot{\Gamma_1}
\xrightarrow{g}
\denot{C}
\end{array}
\]
\[
\begin{array}{l}
\begin{array}{ll}
  \denot{\bang\Delta,\Gamma_1\entail V:\nbang{n}{A}}_\Theta^v
= \denot{\bang\Delta}\tensor\denot{\Gamma_1}
  \xrightarrow{f} L^n\denot{A}
&\qquad
  \denot{\bang\Delta,\Gamma_2\entail W:\nbang{n}{B}}_\Theta^v
= \denot{\bang\Delta}\tensor\denot{\Gamma_2}
  \xrightarrow{g} L^n\denot{B}
\end{array}
\\
\hline
\denot{\bang\Delta,\Gamma_1,\Gamma_2\entail 
  \prodterm{V,W}^n:\nbangp{n}{A\tensor B}}_\Theta^v
=
\denot{\bang\Delta}\tensor\denot{\Gamma_1}\tensor\denot{\Gamma_2}
\xrightarrow{f\tensor_{\bang\Delta} g}
L^n\denot{A}\tensor L^n\denot{B}
\xrightarrow{\cohmap{L^n}_{A,B}}
L^n(\denot{A}\tensor\denot{B})
\end{array}
\]
\vspace{-1.5ex}
Interpretation of computations:
First, if $U$ is a core value, $\denot{\Delta\entail U:A}_\Theta^c =
\denot{\Delta\entail U:A}_\Theta^v;\unit_A$.
\[
\begin{array}{l}
\begin{array}{ll}
  \denot{\bang\Delta,\Gamma_1\entail M:A\loli B}_\Theta^c
= \denot{\bang\Delta}\tensor\denot{\Gamma_1}
  \xrightarrow{f} T(\denot{A}\loli\denot{B})
&\qquad
  \denot{\bang\Delta,\Gamma_2\entail N:A}_\Theta^c
= \denot{\bang\Delta}\tensor\denot{\Gamma_2}
  \xrightarrow{g} T(\denot{A})
\end{array}
\\
\hline
 \denot{\bang\Delta,\Gamma_1,\Gamma_2\entail MN:B}_\Theta^c =
 \denot{\bang\Delta}{\tensor}\denot{\Gamma_1}{\tensor}\denot{\Gamma_2}
 \xrightarrow{f\tensor_{\bang\Delta}g}
  T(\denot{A}{\loli}\denot{B}){\tensor} T(\denot{A})
 \xrightarrow{\Strength_1}
  T((\denot{A}{\loli}\denot{B}){\tensor}\denot{A})
 \xrightarrow{\appnt_{A,B}^*}
  T(\denot{B})
\end{array}
\]
\[
\begin{array}{l}
\begin{array}{ll}
  \denot{\bang\Delta,\Gamma_1
  \entail M:\nbangp{n}{A_1\tensor A_2}}_\Theta^c
&= \denot{\bang\Delta}\tensor\denot{\Gamma_1}
  \xrightarrow{f} TL^n(\denot{A_1}\tensor\denot{A_2})
\\
  \denot{\bang\Delta,\Gamma_2,
    x:\nbang{n}{A_1},y:\nbang{n}{A_2}\entail N:C}_\Theta^v
&= \denot{\bang\Delta}\tensor\denot{\Gamma_2}\tensor
  L^n\denot{A_1}\tensor L^n\denot{A_2}
  \xrightarrow{g} T(\denot{C})
\end{array}
\\
\hline
  \denot{\bang\Delta,\Gamma_2,\Gamma_1\entail
  \nletprodterm{x,y}{n}{M}{N}:\nbang{n}{C}}_\Theta^c
= \denot{\bang\Delta}\tensor\denot{\Gamma_2}\tensor\denot{\Gamma_1}
 \xrightarrow{\id\tensor_{\bang\Delta} f}
  \denot{\bang\Delta}\tensor\denot{\Gamma_1}\tensor 
  TL^n(\denot{A_1}\tensor\denot{A_2})
\\\qquad\xrightarrow{t;
  T\left(\id\tensor\left(\cohmap{L^n}\right)^{-1}{}\right)}
  T(\denot{\bang\Delta}\tensor\denot{\Gamma_1}
    \tensor L^n\denot{A_1}\tensor L^n\denot{A_2})
  \xrightarrow{g^*}T\denot{C}
\end{array}
\]
\[
\begin{array}{l}
\begin{array}{ll}
  \denot{\bang\Delta,\Gamma_2\entail M:\putype}_\Theta^c
= \denot{\bang\Delta}\tensor\denot{\Gamma_2}
  \xrightarrow{f} T(\tensunit)
&\qquad
  \denot{\bang\Delta,\Gamma_1\entail N:C}_\Theta^c
= \denot{\bang\Delta}\tensor\denot{\Gamma_1}
  \xrightarrow{g} T(\denot{C})
\end{array}
\\
\hline
\denot{\bang\Delta,\Gamma_1,\Gamma_2\entail
  \letunitterm{M}{N}:C}_\Theta^c
 =
\denot{\bang\Delta}\tensor\denot{\Gamma_1}\tensor\denot{\Gamma_2}
\xrightarrow{\id\tensor_{\bang\Delta} f}
\denot{\bang\Delta}\tensor\denot{\Gamma_1}\tensor T(\tensunit)
\xrightarrow{t;g^*}
T(\denot{C})
\end{array}
\]
\[
\begin{array}{l}
\begin{array}{ll}
  \denot{\bang\Delta,\Gamma_1\entail M:\nbang{n}{A}}_\Theta^c
= \denot{\bang\Delta}\tensor\denot{\Gamma_1}
  \xrightarrow{f} TL^n\denot{A}
&\qquad
  \denot{\bang\Delta,\Gamma_2\entail N:\nbang{n}{B}}_\Theta^c
= \denot{\bang\Delta}\tensor\denot{\Gamma_2}
  \xrightarrow{g} TL^n\denot{B}
\end{array}
\\
\hline
\denot{\bang\Delta,\Gamma_1,\Gamma_2{\entail} 
  \prodterm{M,N}^n:\nbangp{n}{A{\tensor}B}}_\Theta^c =
\denot{\bang\Delta}{\tensor}\denot{\Gamma_1}{\tensor}\denot{\Gamma_2}
\xrightarrow{f{\tensor_{{\bang\Delta}}}g}
TL^n\denot{A}{\tensor}TL^n\denot{B}
\xrightarrow{\Psi_1;T\cohmap{L^n}_{A,B}}
TL^n(\denot{A}{\tensor}\denot{B})
\end{array}
\]
\end{minipage}}
\tablenl
\mycaption{Interpretation of values and computations.}
\label{table:modelvaluejudg}
\label{table:modelcorevaluejudg}
\label{table:modelcompjudg}
\end{table*}
%.......... END modelvaluejudg .......................................

As we already noted in Section~\ref{sec:typjudg}, a valid typing
judgment does not have a unique typing tree {\em per se}. However the
following result holds:

\begin{theorem}
  Given a valid typing judgment with two typing derivations $\pi$
  and $\pi'$, for any interpretation $\Theta$ we have
  $\denot{\pi}^c_\Theta = \denot{\pi'}^c_\Theta$ (and
  $\denot{\pi}^v_\Theta = \denot{\pi'}^v_\Theta$ if the typing
  judgment is a value).
\end{theorem}

\begin{proof}
  The proof is done by showing that given any typing judgment
  $\Delta\entail M:A$ with denotation $f$ one can factor $f$ as
  $\comonoidunit_{\bang\Gamma}\tensor
  \bar{f}$, where $\bar{f}$ is the denotation of $\Delta'\entail M:A$,
  where $\Delta',\bang\Gamma=\Delta$ and $|\Gamma|$ is the set of
  dummy variables.\qed
\end{proof}

\begin{definition}\label{def:typjudgdenot}\rm
  Given a interpretation $\Theta$ of the language in a category
  $\cat{C}$, we define the denotation of a valid typing judgment
  $\Delta\entail M:A$ with typing derivation $\pi$ to be
  $\denot{\Delta\entail M:A}^c_\Theta = \denot{\pi}^c_\Theta$ and
  $\denot{\Delta\entail M:A}^v_\Theta = \denot{\pi}^v_\Theta$ if $M$
  is a value.
\end{definition}

\begin{lemma}\label{lem:valuemonad}
  Suppose that $\Delta\entail V:A$ is a valid typing judgment where
  $V$ is a value. Then $\denot{\Delta\entail
  V:A}^c= \denot{\Delta} \xrightarrow{\denot{\Delta\entail V:A}^v} \denot{A}
  \xrightarrow{\eta_{\denot{A}}} T(\denot{A})$.
\end{lemma}

\begin{proof}
  Proof by induction on $V$, using
  Lemma~\ref{lem:strongmonadmonoidal}, the bifunctoriality of
  $\tensor_{LA}$ and the equations for strong monadicity in
  Definition~\ref{def:strongmonad}.\qed
\end{proof}

%---------------------------------------------------------------------
%---------------- SUBSECTION -----------------------------------------
%---------------------------------------------------------------------
\subsection{Soundness of the denotation}

The axiomatic equivalence and the categorical semantics are two faces
of the same coin.
Indeed, as we will prove in this section, two terms
in the same axiomatic equivalence class have the same
denotation. A corollary is that the indexation of terms does not
influence the denotation. This proves semantically the fact that it is
safe to work with untyped terms. An alternate justification of this
fact is of course the operational semantics, which was given in
{\cite{selinger05lambda}}.

\begin{lemma}\label{lem:denottypecast}
  Suppose $M'=\s{\Delta'\subtype\Delta\entail M:A\subtype A'}$.
  Then
  $\denot{\Delta'\entail M':A'}^c =
  I_{\Delta',\Delta};\denot{\Delta\entail M:A}^c;T(I_{A,A'})$.
  If $M=V$ is a value, from Lemma~\ref{lem:typecastvalue}, 
  $M'=V'$ is a value. Then
  $\denot{\Delta'\entail V':A'}^v =
  I_{\Delta',\Delta};\denot{\Delta\entail V:A}^v;I_{A,A'}$.
\end{lemma}

\begin{proof}
  Proof by structural induction on $M$.\qed
\end{proof}

\begin{lemma}[Substitution]\label{lem:substitution}
  Given two valid typing judgments $\bang\Delta,\Gamma_1,x:A\entail
  M:B$ and $\bang\Delta,\Gamma_2\entail
  V:A$, the typing judgment
  $\bang\Delta,\Gamma_1,\Gamma_2\entail
  M[V/x]:B$ is valid. Let $h$ be
  $\denot{\bang\Delta,\Gamma_1,\Gamma_2\entail M[V/x]:B}^c$
  and $h'$ be $\denot{\bang\Delta,\Gamma_1,\Gamma_2\entail W[V/x]:B}^v$,
  in the case where $M=W$ is a value. Then they are defined by\\
  \resizebox*{4.85in}{!}{
  $\xymatrix@C=2.5cm{
    \denot{\bang\Delta}\tensor\denot{\Gamma_1}\tensor\denot{\Gamma_2}
    \ar[d]^{\Split_{\bang\Delta,\Gamma_1,\Gamma_2}}
    \ar@{.>}[r]^{h}
    & T(\denot{B})\\
    \denot{\bang\Delta}{\tensor}\denot{\Gamma_1}
    {\tensor}\denot{\bang\Delta}{\tensor}\denot{\Gamma_2}
    \ar[r]^{\id{\tensor}\denot{\bang\Delta,\Gamma_2\entail V:A}^v}&
    \denot{\bang\Delta}{\tensor}\denot{\Gamma_1}{\tensor}\denot{A},
    \ar[u]^{\denot{\bang\Delta,\Gamma_1,x:A\entail
    M:B}^c}
  }
  \quad
  \xymatrix@C=2.8cm{
    \denot{\bang\Delta}\tensor\denot{\Gamma_1}\tensor\denot{\Gamma_2}
    \ar[d]^{\Split_{\bang\Delta,\Gamma_1,\Gamma_2}}
    \ar@{.>}[r]^{h'} & \denot{B}\\
    \denot{\bang\Delta}{\tensor}\denot{\Gamma_1}
    {\tensor}\denot{\bang\Delta}{\tensor}\denot{\Gamma_2}
    \ar[r]^{\id{\tensor}\denot{\bang\Delta,\Gamma_2\entail V:A}^v}
    &  \denot{\bang\Delta}{\tensor}\denot{\Gamma_1}{\tensor}\denot{A}.
    \ar[u]^{\denot{\bang\Delta,\Gamma_1,x:A\entail W:B}^v}
  }$}
\end{lemma}

\begin{proof}
  Proof by induction on $M$, using Lemma~\ref{lem:duplicvalue},
  Lemma~\ref{lem:valuemonad} and the naturality of $\Phi$.\qed
\end{proof}

\begin{theorem}\label{the:aximpliesdenot}
  If $\Delta\entail M\eqax M':A$ then
  $\denot{\Delta\entail M:A}^c_\Theta=\denot{\Delta\entail
  M':A}^c_\Theta$ (and $\denot{\Delta\entail
  M:A}^v_\Theta=\denot{\Delta\entail
  M':A}^v_\Theta$ if $M$ is a value) for every interpretation $\Theta$.
\end{theorem}

\begin{proof}
  Proof by induction on $M\eqax M'$, using
  Lemmas~\ref{lem:denottypecast} and~\ref{lem:substitution}.\qed
\end{proof}

\begin{corollary}\label{cor:erasuredenot}
  If $\Erase(M)=\Erase(M')$ and if $\Delta\entail M,M':A$ are valid
  typing judgments, then $\denot{M}^c=\denot{M'}^c$ (and
  $\denot{M}^v=\denot{M'}^v$ if $M$ is a value).
\end{corollary}

\begin{proof}
  Corollary of Theorems~\ref{the:eraseimpliesax}
  and~\ref{the:aximpliesdenot}.\qed
\end{proof}

%---------------------------------------------------------------------
%---------------- SUBSECTION -----------------------------------------
%---------------------------------------------------------------------
\subsection{Completeness}

The category $\YAQ$ being a linear category for duplication, one
can interpret the language in it. This section states that the defined
lambda-calculus is an \define{internal language} of linear categories
for duplication.

Since the category $\YAQ$ is a monoidal category, one can
w.l.o.g. generalize the notion of pairing to finite tensor products of
terms. Then the following results are true:

\begin{lemma}\label{lem:yaqcomplete}
  In $\YAQ$, a valid typing judgment $x_1:A_1,\ldots
  x_n:A_n\entail M:B$ has for computational denotation
  $(t:A_1\tensor\cdots\tensor A_n\entail \letprodterm{x_1,\ldots
  x_n}{t}{\lambdau.M}:\comptype B)$. If $M=V$ is a value, the value
  interpretation is
  $(t:A_1\tensor\cdots\tensor A_n\entail
  \letprodterm{x_1,\ldots x_n}{t}{V}:B)$.
\end{lemma}

\begin{proof}
  Proof by structural induction on $M$ and $V$.\qed
\end{proof}

\begin{theorem}
  In $\YAQ$, $\Theta$ being the identity, one has
  $\denot{x:A\entail M:B}^c_\Theta \eqax (x:A\entail
  \lambda\puterm.M:\putype\loli B)$ and
  $\denot{x:A\entail V:B}^v_\Theta \eqax (x:A\entail V:B)$.
\end{theorem}

\begin{proof}
  Corollary of Lemma~\ref{lem:yaqcomplete}\qed
\end{proof}

%+++++++++++++++++++++++++++++++++++++++++++++++++++++++++++++++++++++
%++++++++++++++++ SECTION ++++++++++++++++++++++++++++++++++++++++++++
%+++++++++++++++++++++++++++++++++++++++++++++++++++++++++++++++++++++
\section{Towards a denotational model of quantum lambda calculus}

As noted in the introduction, this paper is mostly concerned with the
categorical requirements for modeling a generic call-by-value linear
lambda calculus, i.e., its type system (which includes subtyping) and
equational laws. We have not yet specialized the language to a
particular set of built-in operators, for example, those that are
required for quantum computation.

However, since the quantum lambda calculus {\cite{selinger05lambda}}
is the main motivation behind our work, we will comment very briefly
on what additional properties would be required to interpret its
primitives. The quantum lambda calculus is obtained by instantiating
and extending the call-by-value language of this paper with the
following primitive types, constants, and operations:

\begin{center}
  \begin{tabular}{ll}
    Types: & $\bit$, $\qbit$ \\
    Constants: & $0:\bang{\bit}$, $1:\bang{\bit}$\\
    & $\newterm:\bang{(\bit \loli \qbit)}$, $U:\bang{(\qbit^n\loli\qbit^n)}$, $\measureterm:\bang{(\qbit \loli \bang{\bit})}$\\
    Operations: & $\vcenter{
      \infer[({\it if})]
      { \Gamma_1, \Gamma_2, !\Delta \entail \iftermx{P}{M}{N} : A}
      {
        \Gamma_1, \bang{\Delta} \entail P : bit
        &
        \Gamma_2, \bang{\Delta} \entail M : A
        &
        \Gamma_2, \bang{\Delta} \entail N : A
        }}$
  \end{tabular}
\end{center}

Here, $U$ ranges over a set of built-in unitary operations. In the
intended semantics, $\bang{\bit}\iso\bit$, while $\bang{\qbit}$ is
empty. $\newterm$ creates a new qubit, and $\measureterm$ measures a
qubit. 

The denotational semantics of these operations is already
well-understood in the absence of higher-order types. They can all be
interpreted in the category $\Q$ of superoperators from
{\cite{selinger04quantum}}.  The part that is not yet well-understood
is how these features interact with higher-order types. 

In light of our present work, we can conclude that a model of the
quantum lambda calculus consists of a linear category for duplication
$(\cat{C},L,T,\loli)$, such that the associated category of
computations $\cat{C}_T$ contains the category $\Q$ of
{\cite{selinger04quantum}} as a full monoidal subcategory.  To
construct an actual instance of such a model is still an open problem.

%+++++++++++++++++++++++++++++++++++++++++++++++++++++++++++++++++++++
%++++++++++++++++ SECTION ++++++++++++++++++++++++++++++++++++++++++++
%+++++++++++++++++++++++++++++++++++++++++++++++++++++++++++++++++++++
\section{Conclusion}

We have developed a call-by-value, computational lambda-calculus for
manipulating duplicable and non-duplicable data, together with an
axiomatic equivalence relation on typed terms. We use a subtyping
relation in order to have implicit promotion, dereliction, copying and
discarding. 
Then we developed categorical model for the language, inspired by
the work of \cite{bierman93intuitionistic} and \cite{moggi91notions}.
We finally showed that the model is sound and complete with respect to
the axiomatic equivalence.

%\bibliographystyle{splncssrt}
%\bibliography{paper}

\begin{thebibliography}{10}

\bibitem{abramsky93computational}
Abramsky, S.:
\newblock Computational interpretations of linear logic.
\newblock Theoretical Computer Science \textbf{111} (1993)  3--57

\bibitem{abramsky04categorical}
Abramsky, S., Coecke, B.:
\newblock A categorical semantics of quantum protocols.
\newblock In: Proceedings of LICS'04. (2004)  415--425

\bibitem{barendregt84lambda}
Barendregt, H.P.:
\newblock The Lambda-Calculus, its Syntax and Semantics.
\newblock North Holland (1984)

\bibitem{benton94mixed}
Benton, N.:
\newblock A mixed linear and non-linear logic: Proofs, terms and models
  (extended abstract).
\newblock In: Proceedings of CSL'94, Selected Papers. Volume 933 of Lecture
  Notes in Computer Science. (1994)  121--135

\bibitem{benton93term}
Benton, N., Bierman, G., de~Paiva, V.C.V., Hyland, M.:
\newblock A term calculus for intuitionistic linear logic.
\newblock In: Proceedings of TLCA'93. Volume 664 of Lecture Notes in Computer
  Science. (1993)  75--90

\bibitem{benton92linear}
Benton, N., Bierman, G., Hyland, M., de~Paiva, V.C.V.:
\newblock Linear lambda-calculus and categorical models revisited.
\newblock In: Proceedings of CSL'92, Selected Papers. Volume 702 of Lecture
  Notes in Computer Science. (1992)

\bibitem{benton96linear}
Benton, N., Wadler, P.:
\newblock Linear logic, monads and the lambda calculus.
\newblock In: Proceedings of LICS'96. (1996)  420--431

\bibitem{bierman93intuitionistic}
Bierman, G.:
\newblock On Intuitionistic Linear Logic.
\newblock PhD thesis, Computer Science Department, Cambridge University (1993)

\bibitem{coecke04informationflow}
Coecke, B.:
\newblock Quantum information-flow, concretely, abstractly.
\newblock  \cite{qpl04}  57--73

\bibitem{coecke06quantum}
Coecke, B., Pavlovic, D.:
\newblock Quantum measurements without sums.
\newblock In Chen, G., Kauffman, L., Lomonaco, S.J., eds.: Mathematics of
  Quantum Computation and Technology.
\newblock Chapman \& Hall (2007)  559--596

\bibitem{GaySJ:comqp}
Gay, S.J., Nagarajan, R.:
\newblock Communicating quantum processes.
\newblock In: Proceedings of POPL'05, ACM Press (2005)

\bibitem{LalJor}
Lalire, M., Jorrand, P.:
\newblock A process algebraic approach to concurrent and distributed
  computation: operational semantics.
\newblock  \cite{qpl04}  109--126

\bibitem{maclane98categories}
Mac~Lane, S.:
\newblock Categories for the Working Mathematician.
\newblock Springer Verlag (1998)

\bibitem{moggi91notions}
Moggi, E.:
\newblock Notions of computation and monads.
\newblock Information and Computation \textbf{93} (1991)  55--92

\bibitem{schalk04what}
Schalk, A.:
\newblock What is a model for linear logic.
\newblock Manuscript (2004)

\bibitem{Seely89}
Seely, R.A.G.:
\newblock *-autonomous categories and cofree coalgebras.
\newblock Contemporary Mathematics \textbf{92} (1989)

\bibitem{qpl04}
Selinger, P., ed.:
\newblock Proceedings of QPL'04.
\newblock TUCS General Publication No 33, Turku Centre for Computer Science
  (2004)

\bibitem{selinger04quantum}
Selinger, P.:
\newblock Towards a quantum programming language.
\newblock Mathematical Structures in Computer Science \textbf{14} (2004)
  527--586

\bibitem{selinger05lambda}
Selinger, P., Valiron, B.:
\newblock A lambda calculus for quantum computation with classical control.
\newblock Mathematical Structures in Computer Science \textbf{16} (2006)
  527--552

\bibitem{valiron06fully}
Selinger, P., Valiron, B.:
\newblock On a fully abstract model for a quantum linear functional language.
\newblock In: Preliminary proceedings of QPL'06. (2006)  103--115

\bibitem{tonder04lambda}
van Tonder, A.:
\newblock A lambda calculus for quantum computation.
\newblock SIAM Journal of Computing \textbf{33} (2004)  1109--1135

\bibitem{wadler92substitute}
Wadler, P.:
\newblock There's no substitute for linear logic.
\newblock Manuscript, presented at MFPS'92 (1992)

\bibitem{WZ82}
Wootters, W.K., Zurek, W.H.:
\newblock A single quantum cannot be cloned.
\newblock Nature \textbf{299} (1982)  802--803

\end{thebibliography}

\end{document}

%%% Local Variables: 
%%% TeX-master: "paper.tex"
%%% End: 